\documentclass[twocolumn,pra,a4paper,nofootinbib,showpacs,aps,floatfix,superscriptaddress]{revtex4-1}
\usepackage{graphicx}
\usepackage[utf8]{inputenc}
\usepackage{bm}
\usepackage{amsmath}
\usepackage{amssymb}
\usepackage{enumerate}
\usepackage{color}
\def\la{\langle}
\def\ra{\rangle}

\begin{document}
\title{Fast state and trap  rotation of a particle in an anisotropic potential}
\author{I. Lizuain}
\email{ion.lizuain@ehu.eus}
\affiliation{Department of Applied Mathematics, University of the Basque Country UPV/EHU, Donostia-San Sebastian, Spain}
\author{A. Tobalina}
\affiliation{Department of Physical Chemistry, University of the Basque Country UPV/EHU, Apdo. 644, Bilbao, Spain}
\author{A. Rodriguez-Prieto}
\affiliation{Departament of Applied Mathematics, University of the Basque Country UPV/EHU, Bilbao, Spain}
\author{J. G. Muga}
\affiliation{Department of Physical Chemistry, University of the Basque Country UPV/EHU, Apdo. 644, Bilbao, Spain}

\begin{abstract}
We study the dynamics of a quantum or classical particle in  a two-dimensional  rotating anisotropic harmonic potential. 
By a  sequence of  symplectic transformations for constant rotation velocity  we find uncoupled normal generalized coordinates and conjugate momenta
in which the Hamiltonian
takes the form of two independent harmonic oscillators.  
The decomposition into normal-mode dynamics enables us to design fast trap-rotation processes to produce a rotated version of an arbitrary initial state, 
when the two  normal frequencies are commensurate.
\end{abstract}
\maketitle
%
\section{Introduction}
Motivated by existing or developing  
quantum technologies, much work  is currently being devoted to control the motional dynamics of quantum systems. 
Basic operations such as shuttling, expansions/compressions, merging and separation of atom or ion chains, or rotations of the quantum particles are needed to implement interferometers,
quantum information applications, or quantum thermodynamical devices.  
Performing fast operations that do not leave residual excitations is generically of interest not only to save time but  to avoid decoherence as well.  

\begin{figure}
\begin{center}
\includegraphics[width=7cm]{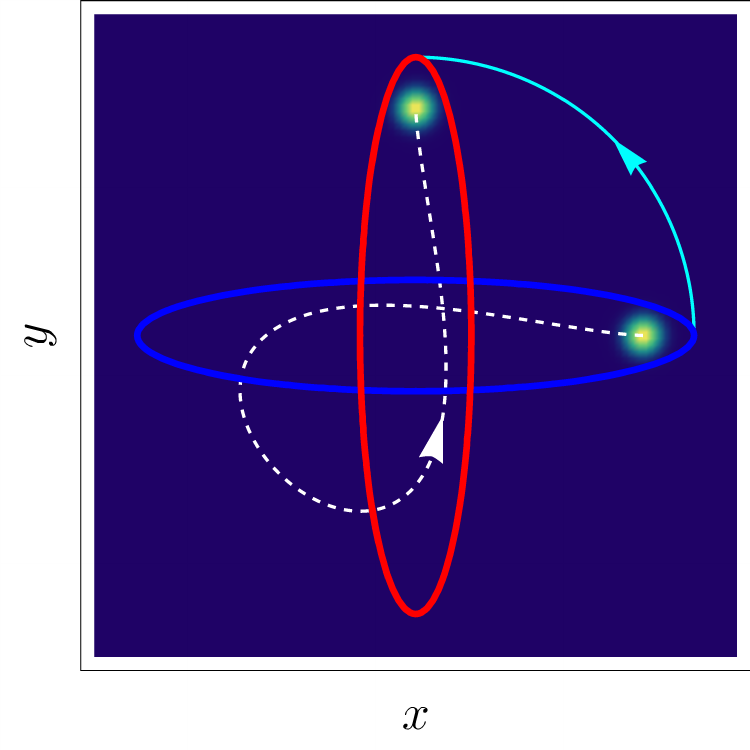}
\caption{(Color online).  Trap rotation in the lab frame (solid line arrow) and particle dynamics (dashed line arrow). The trap is at rest for $t<0$ (horizontal ellipse); then it is rotated by $\pi/2$ from $t=0$ to $t=T$; and finally it remains again at rest for $t>T$ (vertical ellipse). The trap rotation is designed such that the state at time $T$ is the rotated version of the initial state, {\it for all possible -classical or quantum- initial states}.  Just one of them -chosen arbitrarily- is depicted.   The dashed line is the trajectory of the state center  along the trap rotation process.}
\label{lab_frame_fig}
\end{center}
\end{figure}

Shortcuts to adiabaticity (STA) are proposed as a set of efficient techniques to design such operations \cite{Torrontegui2013}. 
For two or more effective dimensions, shortcut design, by inverse engineering the control parameters using invariants of motion, is much facilitated by finding {\it dynamical normal modes} \cite{Palmero2014}. These modes are independent 
harmonic motions  in the regime of small oscillations corresponding in general to time-dependent harmonic oscillators. 
Studies on different operations on  trapped ions \cite{Palmero2013,Palmero2014,Palmero2015a,Palmero2015b,Lu2015,Palmero2016,Palmero2017}
made clear that it is not always possible to find a point transformation\footnote{In a point transformation  the new coordinates   
only depend on the old ones and not on old momenta.} that leads to independent normal modes. The condition that allows to find a point transformation was 
finally  given in \cite{Lizuain2017} for two-dimensional (2D) Hamiltonians: 
the effective potential can be scaled or translated but it should not rotate. 
Thus the rotation of a 2D anisotropic trap is the paradigmatic model in which  such point transformation cannot be made and it was left as an open question if 
more general transformations could be used to speed up the rotation \cite{Lizuain2017}. The inertial effect due to the trap rotation can be formally compensated  
by an effective angular momentum term \cite{Masuda,Lizuain2017} to leave the particle at rest in the rotating frame. This term though may be
difficult to implement, for example if the particle is not charged,  so we consider in this paper that the only manipulation available is the rotation of the trap itself, without any additional force.  
STA for simple 1D-trap rotations, without compensation terms,  were described in \cite{Palmero2016}
but STA for the more realistic 2D anisotropic trap had not been described.    

The goal of this work is to perform a rotation as represented schematically in Fig. 1 in the lab frame of coordinates $x,y$:  The trap is at rest for $t<0$; then it is rotated up to time $T$; and finally it remains again at rest for $t>T$. The trap rotation must be designed such that the state at time $T$ is exactly the rotated version of the initial state at time $t=0$, {\it{for all possible initial states}}. Equivalently, from the point of view of the rotating frame, the objective is to get at time $T$ the same state that was prepared at time $0$, regardless of what that state may be.

Rotations of condensates or of  a few particles are of interest for different reasons, such as reordering chains,
redirecting, squeezing \cite{Palmero2016}, 
or creating artificial magnetism \cite{Fetter2009,Dalibard2011}. Here we treat the simplest case of a single particle in a rotating 2D trap. The operation would be 
instrumental in driving atoms through corners and junctions in a scalable quantum processor \cite{20,21}, and may be regarded as a 
first step towards the more difficult problem of rotating  ion chains \cite{17, 20, 22}, which would facilitate scalability in linear  traps, and be useful to rearrange the chain, e.g. to locate a cooling ion at the right position in the chain \cite{17}. Rotor states have other applications in sensing, metrology, 
and fundamental physics studies \cite{Urban2019}.  
  
The treatment and transformations are done first in a classical setting. However, since we deal with a harmonic anisotropic trap the results can 
be translated into  quantum mechanics rather directly.  
After setting the model in Sec. \ref{pm},  the independent  normal modes will be first defined and characterized by normal frequencies in Sec. \ref{symplectic_diag_sec}. 
Section \ref{fr} analyzes the fast rotations that may be achieved at certain process times for configurations in which the normal frequencies are commensurate. 
The  minimal time is identified, examples are given,  and a stability analysis is carried out. Finally, Sec. \ref{d} discusses some open questions. 
             
%
%


\section{Physical model\label{pm}}
Our starting point is the Hamiltonian of a particle of mass $m$ in a 2D anisotropic harmonic potential with axial (angular) frequencies $\omega_1$ and $\omega_2$, 
which  rotates around the $z$ axis perpendicular to the trap plane by an angle $\theta$ 
with an angular velocity $\dot\theta$, see Fig. \ref{lab_frame_fig}. 
In the rotating frame of coordinates $\{\tilde q_1,\tilde q_2\}$ and momenta $\{\tilde p_1,\tilde p_2\}$ the Hamiltonian is given by, see Appendix \ref{lab_frame_app},
\begin{eqnarray}
\label{Hrf_orig}
 H&=&\frac{\tilde p_1^2}{2m}+\frac{\tilde p_2^2}{2m} +\frac{1}{2}m\omega_1^2 \tilde q_1^2
 +\frac{1}{2}m\omega_2^2\tilde q_2^2
 -\dot\theta L_z,
\end{eqnarray}
where $L_z= \tilde q_1  \tilde p_2- \tilde q_2 \tilde p_1$.
$H$ has the form of two  harmonic oscillators coupled by an angular momentum $L_z$
that accounts for the inertial effects \cite{Goldstein}.
 
By introducing the dimensionless coordinates and momenta
\begin{eqnarray}
 q_j=\sqrt{\frac{m\omega_j}{\hbar}}\tilde q_j,
 \;\;\;
 p_j=\frac{ \tilde p_j}{ \sqrt{m\hbar \omega_j}},
\end{eqnarray}
the Hamiltonian (\ref{Hrf_orig}) can be written ($\hbar=1$ hereafter) as 
\begin{eqnarray}
  \label{Hrf}
  H&=&\frac{\omega_1}{2}\left( p_1^2+q_1^2\right)   + \frac{\omega_2}{2}\left( p_2^2+q_2^2\right)
  \nonumber\\
  &-&\dot\theta\left(\frac{1}{\eta}q_1p_2-\eta q_2p_1\right),
 \end{eqnarray}
where $\eta=\sqrt{\omega_1/\omega_2}$.
 
This rotating frame Hamiltonian depends only on the angular velocity $\dot\theta$ as a control parameter.
We shall consider throughout the work a constant rotation velocity, i. e., a linear-in-time  angle $\theta(t)=\dot\theta t$
from $t=0$ to $t=T$. Thus the Hamiltonian in the rotating frame is 
time independent during the rotation. 

The Hamiltonian (\ref{Hrf}) can be written in compact matrix representation as the quadratic form
\begin{equation}
\label{Hrf_matrixform}
H= v^T A  v,
\end{equation}
where $ v^T=(q_1,q_2,p_1,p_2)$ and $A$ is the symmetric $4\times 4$ matrix
\begin{eqnarray}
\label{A_rot}
  A&=&\frac{1}{2}\left(  \begin{matrix} \omega_1&0&0&-\frac{\dot\theta}{\eta}\\ 0&\omega_2&\eta\dot\theta&0\\0&\eta\dot\theta&\omega_1&0\\-\frac{\dot\theta}{\eta}&0&0&\omega_2 \end{matrix}\right).
\end{eqnarray}
Our first goal is to find a transformation to a frame  in which  the corresponding effective Hamiltonian is uncoupled in both coordinates and momenta, or, using the four-dimensional matrix formalism, it 
is characterized by a diagonal  matrix. To do so we will use the symplectic approach to canonical transformations.
%
%
%
%
\section{Symplectic diagonalization}
\label{symplectic_diag_sec}
In the $4\times4$ matrix representation presented above, a canonical transformation
will be defined by the transformation $ v = S  V $ to a new set of canonical coordinates $ V ^T=(Q_1,Q_2,P_1,P_2)$
provided $S$ is a $4\times 4$ symplectic matrix.
A symplectic matrix $S$ satisfies  $S^T J S=J$, where $J$ is the skew-symmetric matrix \cite{Goldstein}
\begin{eqnarray}
\label{symplectic_Jmatrix}
 J&=&\left(\begin{matrix} 0&0&1&0\\0&0&0&1\\-1&0&0&0\\0&-1&0&0\end{matrix}\right).
\end{eqnarray}
Note that its  inverse is simply $J^{-1}=J^T=-J$. As well, $S^{-1}=J^{-1}S^TJ$.  
$4\times 4$ real symplectic matrices form the ten-dimensional symplectic group $Sp(4,\mathbb{R})$ \cite{simon1999}.
Applying a symplectic (i. e., canonical) transformation to $H$ amounts to rewrite it as  
\begin{eqnarray}
\label{H_transformation}
  H= v^T A  v= V ^T \left(S^T A S\right)   V .
 \end{eqnarray}
%
%
Given the matrix $A$ (\ref{A_rot}) we want to find a symplectic matrix $S\in Sp(4,\mathbb{R})$ so that $S^T A S$ is a diagonal
matrix. 
Such a diagonalizing symplectic matrix $S$ will always exist 
as long as $A$ is a positive definite matrix. This result is known as  Williamson's Theorem
\cite{Williamson1936,Pirandola2009,Gosson}.
The positivity of $A$ imposes an upper bound for the allowed rotation velocity  in order to end up with an uncoupled effective Hamiltonian. 
In particular, the rotation velocity must satisfy
\begin{eqnarray}
\label{williamson_condition}
 \dot\theta<\textrm{min}(\omega_1,\omega_2).
\end{eqnarray}
For simplicty, and without loss of generality, we will consider  $\omega_1<\omega_2$ (or $0<\eta<1$) throughout this work. 
Therefore, the three (angular) frequencies  in our model satisfy the conditions 
\begin{equation}
\label{3freq_cond}
 \dot\theta<\omega_1<\omega_2.
\end{equation}
%
%
%
 \subsection{Constructing the $S$ matrix}
 \label{transformations_sequence}
We will construct the $S$ matrix after a four-step sequence of symplectic transformations.


 {\it i)} The first transformation  brings the matrix  $A$  (\ref{A_rot}) to a block diagonal form. 
 This is achieved by the symplectic matrix 
 \begin{eqnarray}
  S_0=\left(\begin{matrix} 0 & 0 & -1 & 0 \\ 0 & 1 & 0 & 0 \\ 1 & 0 & 0 & 0 \\ 0 & 0 & 0 & 1 \end{matrix}\right),
 \end{eqnarray}
which leads to 
 \begin{eqnarray}
A_1&=&S_0^T A S_0 =\frac{1}{2}\left(\begin{matrix} \omega_1 & \eta  \dot\theta & 0 & 0 \\ \eta  \dot\theta & \omega_2 & 0 & 0 \\ 0 & 0 & \omega_1 & \frac{\dot\theta}{\eta } \\ 0 & 0 & \frac{\dot\theta}{\eta } & \omega_2\end{matrix}\right).
\end{eqnarray}
This transformation is not  
a point  transformation since $S_0$ mixes coordinates and momenta  as already noted in \cite{Manko1996}. 

{\it ii)} The second transformation  diagonalizes one of the two blocks in $A_1$. 
We choose the lower one in this case 
(the ``momenta block'')\footnote{If $\omega_1>\omega_2$ had been assumed,
at this point the upper block should be diagonalized instead of the lower one}. 
This transformation is performed  by the symplectic matrix
 \begin{eqnarray}
   S_1&=&\left(\begin{matrix} 1 & \frac{\dot\theta}{\sqrt{\omega_1\omega_2}} & 0 & 0 \\ 0 & 1 & 0 & 0 \\ 0 & 0 & 1 & 0 \\ 0 & 0 & -\frac{\dot\theta}{\sqrt{\omega_1\omega_2}} & 1  \end{matrix}\right)
 \end{eqnarray}
and leads to
 \begin{eqnarray}
A_2&=&S_1^T A_1 S_1
=\frac{1}{2}\left(\begin{matrix} \omega_1 & 2\eta \dot\theta& 0 & 0 \\2\eta \dot\theta& \frac{3 \dot\theta^2+\omega_2^2}{\omega_2} & 0 & 0 \\ 0 & 0 & \frac{\omega_1^2-\dot\theta^2}{\omega_1} & 0 \\ 0 & 0 & 0 & \omega_2\end{matrix}\right).
 \end{eqnarray}
 %

{\it iii)} The third step  transforms the block that it is already diagonal (the lower block in our case) 
into the identity. This is achieved by the symplectic matrix
 \begin{eqnarray}
S_2&=&\left(
\begin{matrix} \sqrt{\frac{\omega_1^2-\dot\theta ^2}{\omega_1}} & 0 & 0 & 0 \\
 0 & \sqrt{\omega_2} & 0 & 0 \\
 0 & 0 & \sqrt{\frac{\omega_1}{\omega_1^2-\dot\theta ^2}} & 0 \\
 0 & 0 & 0 & \frac{1}{\sqrt{\omega_2}}
 \end{matrix}\right),
 \end{eqnarray}
which transforms $A_2$ into
\begin{eqnarray}
 A_3&=&S_2^T A_2 S_2\nonumber\\
&=&\frac{1}{2}\left(\begin{matrix}  \omega_1^2-\dot\theta^2 & 2\dot\theta \sqrt{\omega_1^2-\dot\theta^2} & 0 & 0 \\
2\dot\theta \sqrt{\omega_1^2-\dot\theta^2} &  3\dot\theta^2+\omega_2^2 & 0 & 0 \\
 0 & 0 & 1 & 0 \\ 0 & 0 & 0 & 1 \end{matrix}\right).
\end{eqnarray}
The transformation requires $\omega_1>\dot\theta$, which is consistent with Eq. (\ref{3freq_cond}).

\begin{figure}
\begin{center}
\includegraphics[width=8cm]{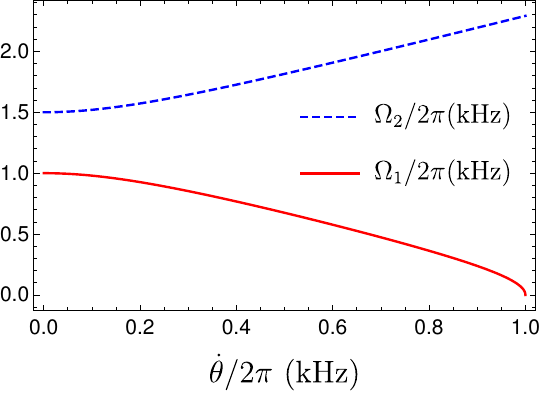}
\caption{(Color online) Normal mode frequencies $\Omega_1$ (red solid) and $\Omega_2$ (blue dashed) as a function of the rotation angular velocity
for axial frequencies $\omega_1=2\pi\times 1$ kHz and $\omega_2=1.5\, \omega_1$.
There is a maximum allowed $\dot\theta$, when one of the normal mode frequencies becomes complex $\dot\theta_{max}=\omega_1$, see Eq.  (\ref{3freq_cond}).
For a non-rotating trap  ($\dot\theta=0$), these frequencies are simply the axial frequencies $\omega_{1,2}$.
}
\label{nmfreqs_fig}
\end{center}
\end{figure}

 {\it iv)} Finally, a (formal) rotation of an angle $\alpha$ brings the upper block to a diagonal form, leaving the lower block unaltered, 
\begin{eqnarray}
 S_3&=&\left(\begin{matrix}
 \cos\alpha & -\sin\alpha & 0 & 0 \\
 \sin\alpha & \cos\alpha & 0 & 0 \\
 0 & 0 & \cos\alpha & -\sin\alpha \\
 0 & 0 & \sin\alpha & \cos\alpha 
\end{matrix}\right),
\end{eqnarray}
 with the angle of rotation $\alpha$  given by
\begin{eqnarray}
 \tan2\alpha=\frac{4\dot\theta\sqrt{\omega_1^2-\dot\theta^2}}{\omega_1^2-\omega_2^2-4\dot\theta^2}.
\end{eqnarray}
This last transformation, leads  to our objective, a diagonal matrix
 \begin{eqnarray}
  A_4&=&S_3^T A_3 S_3= \frac{1}{2}\left(  \begin{matrix}\Omega_1^ 2&0&0&0\\0&\Omega_2^2&0&0\\ 0 & 0 & 1 & 0 \\ 0 & 0 & 0 & 1 \end{matrix}\right),
  \label{A4}
 \end{eqnarray}
where the $\Omega_{1,2}$ are the normal mode frequencies with  squares
\begin{eqnarray}
  \Omega_{1}^2&=&\dot\theta ^2+\frac{\omega_1^2+\omega_2^2}{2}-\frac{1}{2}\sqrt{8 \dot\theta ^2\! \left(\omega_1^2\!+\!\omega_2^2\right)\!+\!\left(\omega_1^2-\omega_2^2\right)^2},  \nonumber\\
  \Omega_{2}^2&=&\dot\theta ^2+\frac{\omega_1^2+\omega_2^2}{2}+\frac{1}{2}\sqrt{8 \dot\theta ^2\! \left(\omega_1^2\!+\!\omega_2^2\right)\!+\!\left(\omega_1^2-\omega_2^2\right)^2},\nonumber\\
\label{normal_mode_freqs}
\end{eqnarray}
see a plot of these frequencies as a function of $\dot\theta$ in Fig. \ref{nmfreqs_fig}.
These eigenfrequencies have been found before by Bialynicki-Birula using a different approach \cite{BB97}. 
Our four-step method is sketched  in  \cite{Manko1996}, although the eigenfrequencies and explicit transformations 
were not given there.

%
%
%
%
%
 \subsection{Uncoupled  Hamiltonian and normal modes} 
 After the sequence of four different transformations, the symplectic matrix we were looking for can be written as
 (the product of symplectic matrices is symplectic)
\begin{eqnarray}
\label{Stot}
 S&=&S_0S_1S_2S_3.
\end{eqnarray}
$S$ diagonalizes the initial $A$ matrix by the relation $A_4=S^T A S$ and 
relates old coordinates and momenta  in the rotating-frame and new coordinates and momenta in the transformed frame by the transformation $ v= S  V $ or
\begin{eqnarray}
\label{class_transf}
 \left(\begin{matrix} q_1\\q_2\\p_1\\p_2\end{matrix}\right)=S\left(\begin{matrix} Q_1\\Q_2\\P_1\\P_2\end{matrix}\right).
\end{eqnarray}
By inverting this relation, we can give explicit expressions for the new frame coordinates and momenta in terms of the original ones,
\begin{eqnarray}
\label{Qq_transf}
\left \{ \begin{matrix}
 Q_1&=&q_2\frac{ \sqrt{\delta } \sin \alpha-\dot\theta  \cos\alpha}{\sqrt{\delta  \omega_2}}   +   p_1   \sqrt{\frac{\omega_1}{\delta }}  \cos \alpha\\
 Q_2&=&q_2 \frac{\sqrt{\delta } \cos \alpha+\dot\theta  \sin \alpha}{\sqrt{\delta  \omega_2}}-p_1 \sqrt{\frac{\omega_1}{\delta }}  \sin \alpha \\
 P_1&=&-q_1 \frac{ \sqrt{\delta } \cos \alpha+\dot\theta  \sin \alpha}{\sqrt{\omega_1}}  + p_2 \sqrt{\omega_2} \sin \alpha\\
 P_2&=&q_1\frac{\sqrt{\delta } \sin \alpha-\dot\theta  \cos \alpha}{\sqrt{\omega_1}} + p_2 \sqrt{\omega_2} \cos \alpha
 \end{matrix} \right.,
\end{eqnarray}
with $\delta=\omega_1^2-\dot\theta^2$, which makes clear that this is not a point transformation. 
%
The Hamiltonian written in normal-mode coordinates and momenta takes the simple form of two independent
harmonic oscillators with  normal frequencies $\Omega_{1,2}$,
\begin{eqnarray}
\label{kamiltonian}
H&=& v^T A  v= V ^T S^T A S  V = V ^T A_4  V \nonumber\\
 &=&\frac{1}{2} \left(P_1^2+P_2^2+\Omega_1^2 Q_1^2+\Omega_2^2 Q_2^2\right).
\end{eqnarray}
As discussed in Appendix \ref{unitary_transf_app}, these transformations are identical for a quantum Hamiltonian 
and can be related to quantum unitary transformations. 
Therefore, the Hamiltonian (\ref{kamiltonian}) can be quantized by substituting the generalized  coordinate and momenta by the corresponding operators. 
Since the Hamiltonian is quadratic, we may equivalently rely on a phase-space description 
of the quantum state dynamics in Wigner representation. The dynamics of the  Wigner function is governed  by a classical Liouville equation; equivalently, a phase-space point is driven by classical Hamiltonian dynamics.

\section{Fast rotations\label{fr}}
%
\subsection{Commensurate anisotropic oscillator}
The time evolution generated by the Hamiltonian (\ref{kamiltonian}) is governed by two independent harmonic oscillators. In this frame, 
the corresponding classical trajectories will be given by   Lissajous-like orbits, that 
will only be closed when the ratio between the $\Omega_{1,2}$ frequencies is a rational number, i. e., when 
they are commensurate.
Let us suppose that $n_{1,2}$ are two integers ($n_1<n_2$). Then, if the condition
\begin{eqnarray}
\label{conmensurable}
 \frac{\Omega_2}{\Omega_1}=\frac{n_2}{n_1}
\end{eqnarray}
is satisfied, the full period of the dynamics is  given by 
\begin{eqnarray}
\label{T_period}
 T=\frac{2\pi n_1}{\Omega_1}=\frac{2\pi n_2}{\Omega_2}.
\end{eqnarray}
If a rotation  is performed in a time  $T$, the system will end up in the same initial state in the rotating frame:
the first oscillator performs $n_1$ oscillations, and the second one  $n_2$ full oscillations.

To perform a rotation of an angle $\theta_f=\dot\theta T$ (assuming an initial angle $\theta_i=0$)
at a constant angular velocity $\dot\theta$ in time  $T$, the above relation may be written as
\begin{eqnarray}
\label{theta_final}
 T=\frac{\theta_f}{\dot\theta}=\frac{2\pi n_1 }{\Omega_1(\dot\theta,\omega_1,\omega_2)}=\frac{2\pi n_2}{\Omega_2(\dot\theta,\omega_1,\omega_2)}.
\end{eqnarray}
For some fixed values of $\theta_f$, $n_1$ and $n_2$, these equalities do not have a unique solution since there are two equations but three different parameters
(rotation velocity $\dot\theta$, and frequencies $\omega_1$ and $\omega_2$). 
Using Eq. (\ref{theta_final}) we may write two of the frequencies in terms of a third one, for instance
\begin{eqnarray}
 \omega_1&=&\kappa_-\dot\theta,
 \nonumber\\
 \omega_2&=&\kappa_+\dot\theta,
\label{k_ratios}
\end{eqnarray}
where 
%
\begin{eqnarray}
\label{kappapm}
 \kappa_{\pm}=\left(\!-1\!+\!\frac{2 \pi ^2 \delta_+}{\theta_f^2}\!\pm\!\frac{2 \sqrt{\pi ^4 \delta_-^2-2 \pi ^2 \delta_+ \theta_f^2+\theta_f^4}}{\theta_f^2}\right)^{\!\!\!\!1/2}
\end{eqnarray}
with $\delta_\pm=n_1^2\pm n_2^2$. 
%
Once one of the frequencies is fixed, the remaining two will be 
determined by Eq. (\ref{k_ratios}). In the following, the value of the smallest axial frequency $\omega_1$
will be fixed, but a similar analysis could be done if any of the two remaining
ones is fixed.\footnote{Also of interest is the setting where $\omega_1$ and $\omega_2$ are given, i.e., we do not assume that their values  can be controlled. Then $\dot{\theta}$ for different $n_{1}, n_2$  should be adjusted  to satisfy the last equality in Eq. (\ref{theta_final}). Since $T$ is fixed by the last  two ratios in Eq. (\ref{theta_final}), only a set of discrete values of  $\theta_f$ are allowed in this scenario.}

For a given value of $\omega_1$, relation (\ref{k_ratios}) determines $\dot\theta$ and $\omega_2$,
and using Eq. (\ref{theta_final}) the time duration of the rotation operation is
\begin{eqnarray}
\label{T_duration}
 T_{n_1,n_2}&=&\frac{\kappa_-\theta_f}{\omega_1},
\end{eqnarray}
which, for some fixed values of $\omega_1$ and $\theta_f$, is just a function of the integers $n_1$ and $n_2$.
See some numerical values of $T_{1,2}$ for a $\pi/2$ rotation  in Table \ref{data_table}.
%
\begin{table}
\centering
\begin{tabular}{|l| l|l|l|}
\hline
$\omega_1$& $\omega_2$ &  $\dot\theta$  & $T_{1,2}=\theta_f/\dot\theta$\\
\hline\hline
$2\pi\times 1 $ kHz&$2\pi\times 1.79$ kHz&$2\pi\times 0.23$ kHz&$1.08$ ms\\
$2\pi\times 2 $ kHz&$2\pi\times 3.59$ kHz&$2\pi\times 0.46$ kHz&$0.54$ ms\\
$2\pi\times 5 $ kHz&$2\pi\times 8.96$ kHz&$2\pi\times 1.16$ kHz&$0.22$ ms\\
$2\pi\times 10 $ kHz&$2\pi\times 17.93$ kHz&$2\pi\times 2.32$ kHz&$0.11$ ms\\
\hline
\end{tabular}
\caption{Some numerical values of the trapping frequencies $\omega_1$ and $\omega_2$, rotation angular velocity $\dot\theta$ and time duration $T$ of the rotation operation 
calculated according to Eqs.  (\ref{theta_final}-\ref{T_duration}). 
$\theta_f=\pi/2$, $n_1=1$, and $n_2=2$.}
\label{data_table}
\end{table}
%
%
%
%
\subsection{Fast rotations}
\label{fast_rot_sec}
\begin{figure}[t!]
\begin{center}
\includegraphics[width=8cm]{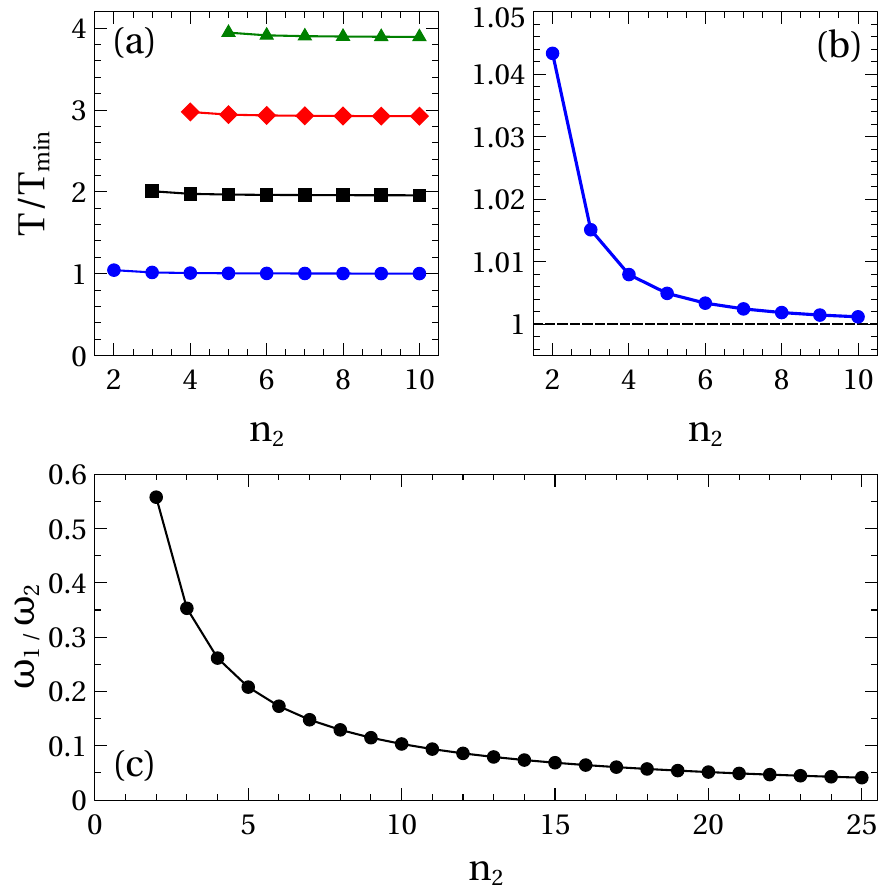}
\caption{(Color online) 
(a) Time $T$ to perform a rotation of $\pi/2$ without final excitation as a function of $n_2$  for different values of $n_1$: $n_1=1$ blue circles, $n_1=2$ black squares,
$n_1=3$ red diamonds and $n_1=4$ green triangles.
The fastest possible rotation corresponds to $n_1=1$ and $n_2\to \infty$.
(b) Closer look at the $n_1=1$ series.
(c) Ratio between axial frequencies $\omega_1/\omega_2=\kappa_-/\kappa_+$ for $n_1=1$. As $n_2$ increases the trap gets narrower.
The fastest possible rotation, at the  $n_2\to\infty$ limit, occurs for an infinitely narrow trap $\omega_1\ll\omega_2$.}
\label{commensurate_fig}
\end{center}
\end{figure}

In principle, the values of $n_1$ and $n_2$ can be chosen arbitrarily as long as $n_1<n_2$:
The time duration of a given rotation (for given $\theta_f$ and $\omega_1$) will be completely determined by the factor $\kappa_-$.
As it is shown in Fig. \ref{commensurate_fig}a-b, the fastest possible rotation (minimum value of $\kappa_-$) is found  with the values $n_1=1$ and $n_2\rightarrow \infty$.
This means that the minimum rotation time $T_{\rm{min}}$ corresponds to a single oscillation of the first (slow) normal mode oscillator and to infinitely many oscillations of 
the second one,
%
\begin{eqnarray}
 T_{\rm{min}}=T_{1,\infty}=\sqrt{\frac{\theta_f^2+4\pi^2}{\omega_1^2}}.
\end{eqnarray}
This minimal time corresponds to  the $\omega_1\ll\omega_2$ limit (i. e., an infinitely narrow trap) as shown in Fig. \ref{commensurate_fig}c.

Of course this limit is an idealization and in practice $\omega_2$ will have some maximal value. 
To illustrate features of a generic case ($n_2\ne \infty$) we choose $n_1=1$ and $n_2=2$ in numerical calculations. 

%

\subsection{Time evolution of states and observables}
In the  reference system of the normal modes, $\{Q_1,Q_2\}$,   a general  wave function  takes the form
\begin{eqnarray}
\psi(Q_1,Q_2,t)&=&\sum_{j=0}^\infty\sum_{j'=0}^\infty c_{jj'} \phi^{(1)}_j(Q_1)e^{-i\Omega_1(j+\frac{1}{2})t}
\nonumber\\
&\times&\phi^{(2)}_{j'} (Q_2)e^{-i\Omega_2(j'+\frac{1}{2})t},
\label{exact}
\end{eqnarray}
where the $c_{ij}$ are constant coefficients set by the initial conditions and $\phi^{(1,2)}_j(Q_{1,2})$ are the usual stationary eigenfunctions of the harmonic oscillators.
If the rotation continues indefinitely,  at a time $t+T$ with $T$ given in Eq. (\ref{T_period}), one gets
\begin{eqnarray}
\psi(Q_1,Q_2,t+T)&=&
\sum_{j=0}^\infty\sum_{j'=0}^\infty c_{jj'} \phi^{(1)}_j(Q_1)e^{-i\Omega_1(j+\frac{1}{2})\left(t+\frac{2\pi n_1}{\Omega_1}\right)}
\nonumber\\
&\times&\phi^{(2)}_{j'} (Q_2)e^{-i\Omega_2(j'+\frac{1}{2})\left(t+\frac{2\pi n_2}{\Omega_2}\right)}
\nonumber\\
&=&(-1)^{n_1+n_2}  \psi(Q_1,Q_2,t),
\end{eqnarray}
i. e., the wave function one period $T$ earlier, with an overall phase that depends on $n_1+n_2$.
The quantum system is said to experience ``exact revivals'' at intervals of $T$  \cite{Razi2000}.
Here we are interested in setting $t=0$ and the corresponding revival at $T$.  

We may use Eq. (\ref{exact}) to find the wavefunction in the rotating frame. To perform the transformation
back and forth between the rotating (coupled)  and  decoupled frames, mixing positions and momenta,  a good strategy is to work in a mixed rotating-frame  representation and use $\la Q'_1,Q'_2|p_1, q_2\ra=\delta[Q'_1-Q_1(p_1,q_2)] \delta[Q'_2-Q_2(p_1,q_2)]$.    
Of course the dynamics may also be solved entirely in the rotating frame  
%
%
%
by numerical integration in a finite basis,\footnote{Specifically in a truncated Fock space  for the interaction-free 
part (two harmonic oscillators) enlarged until converge is achieved.}   
or using the Wigner representation by solving individual trajectories or a system of equations for the moments.

{\subsubsection{Periodic orbits in the rotating frame}}
In the normal-mode frame, the classical trajectories or corresponding center of a wavepacket describe closed Lissajous orbits for commensurate normal frequencies.  
In the rotating frame we find also  corresponding closed  orbits. 
%

\begin{figure}
\begin{center}
\includegraphics[width=7.5cm]{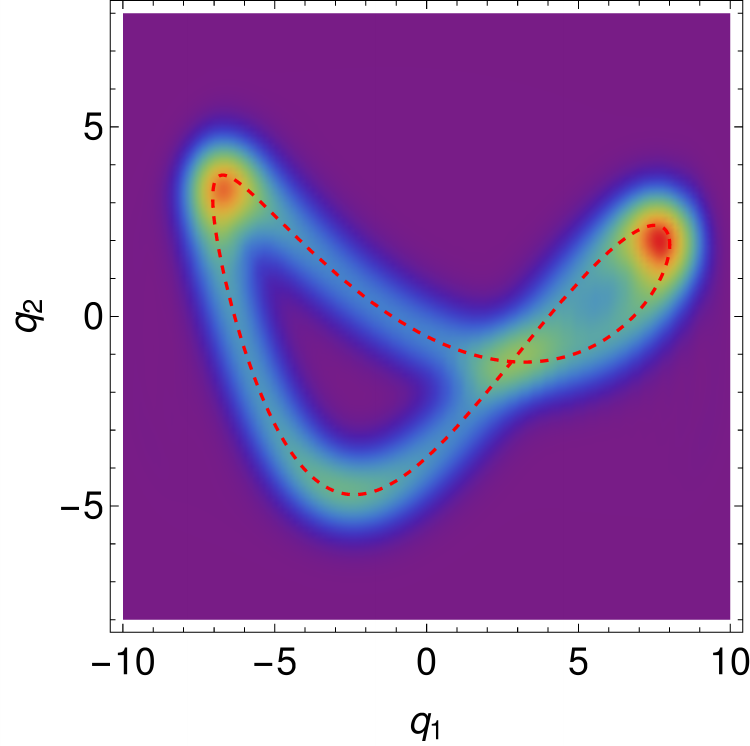}
\caption{(Color online) Wavepacket track ${\cal P}(q_1,q_2)$ of the two mode coherent state $|\alpha_1,\alpha_2\rangle$ 
for the values (initial conditions) $\alpha_1=8/\sqrt2$ and $\alpha_2=2/\sqrt2$ during a rotation of an angle of $\pi/2$. 
Red dashed line: corresponding classical trajectory  with initial conditions
$q_1(0)=\sqrt2 |\alpha_1|=8$, $q_2(0)=\sqrt2 |\alpha_2|=2$ and $p_1(0)=p_2(0)=0$.
The trap and rotation parameters are those in the first row of Table \ref{data_table}.
Dimensionless spatial coordinates $q_1$ and $q_2$ have 
been used as explained in the text.}
\label{trajectory_fig}
\end{center}
\end{figure}

To visualize them let us 
suppose  that the system is initially in the two-mode coherent state $|\psi(0)\rangle =|\alpha_1,\alpha_2\rangle$. 
The state $|\alpha_1,\alpha_2\rangle$ may be expanded in terms of the harmonic oscillators with frequencies $\omega_{1,2}$, 
\begin{eqnarray}
 |\alpha_1,\alpha_2\rangle&=&e^{-\frac{1}{2}\left(|\alpha_1|^2+|\alpha_2|^2\right)}\sum_{n_1,n_2=0}^\infty\frac{\alpha_1^{n_1}\alpha_2^{n_2}}{\sqrt{n_1!n_2!}}|n_1,n_2\rangle,\nonumber
\end{eqnarray}
with $\alpha_j=|\alpha_j|e^{i\varphi}$ being  complex quantities characterizing the (square root) of the average excitation number  and phase of the coherent state.
The state-ket  $|n_1,n_2\rangle$ refers to the eigenstate of the 2D harmonic oscillator with a normalized spatial representation 
\begin{eqnarray}
   \langle q_1,q_2|n_1,n_2\rangle&=&\frac{e^{-\frac{q_1^2+q_2^2}{2}}H_{n_1}(q_1) H_{n_2}(q_2)}{\sqrt{2^{n_1+n_2}n_1!n_2! \pi}},
\end{eqnarray}
where $H_n(q)$ is the  the $n$th order Hermite polynomial.
The time-evolved two-mode coherent state in coordinate representation will be given by the wave function
\begin{eqnarray}
\langle q_1,q_2|\psi(t)\rangle =\langle q_1,q_2|e^{-i H t} |\alpha_1,\alpha_2\rangle
\end{eqnarray}
By integrating the probability density over a full period $T$,
\begin{eqnarray}
 {\cal P} (q_1,q_2)=\int_0^T |\la q_1, q_2|\psi(t)\ra|^2 dt,
\end{eqnarray}
a track of the wave-packet is found, see Fig. \ref{trajectory_fig}, which is more intense where the motion is slow.
The center of the wave-packet follows the classical closed Lissajous-like orbits, ending in its initial configuration after a full rotation is performed.

\begin{figure}
\begin{center}
\includegraphics[width=8.75cm]{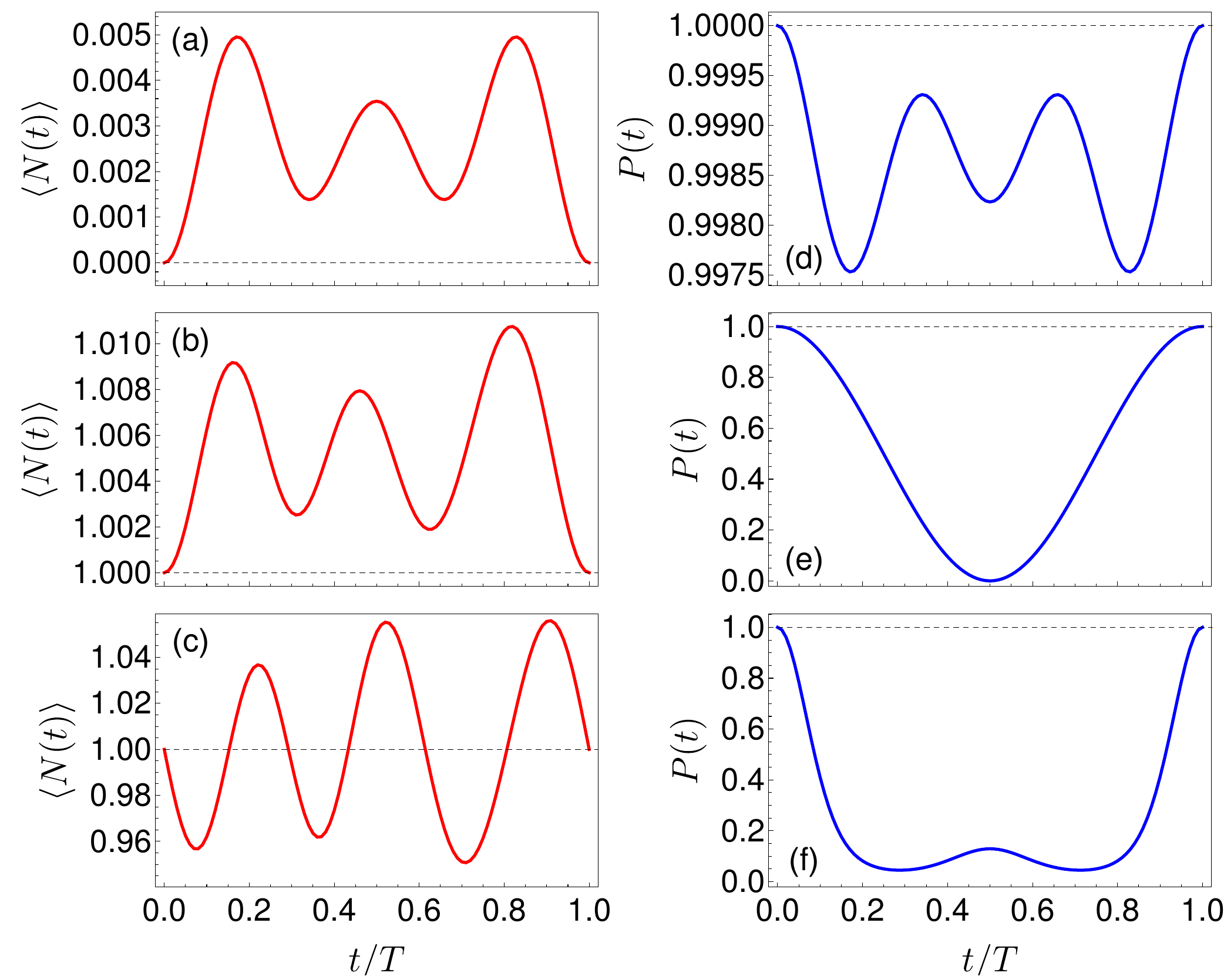}
\caption{(Color online) Time evolution of different observables during a $\theta_f=\pi/2$ rotation. 
In the left column the evolution of the average number of excitations $\langle N(t)\rangle$ is plotted as a function of time, 
while in the second the survival probability $P(t)=|\langle  \psi(t)| \psi(0)\rangle|^2$ of finding the system in its initial state is plotted.
Different initial states are considered for each figure:
In (a) and (d), the  initial state is the ground state of a 2D oscillator  $| v(0)\rangle=|0,0\rangle$ with $\langle N(0)\rangle=0$.
In (b) and (e), the initial state is an entangled state $| v(0)\rangle=\frac{1}{\sqrt2}\left(|0,1\rangle+|1,0\rangle\right)$ with $\langle N(0)\rangle=1$.
In (c) and (f), the initial state is a coherent state $| v(0)\rangle =|\alpha_1,\alpha_2\rangle$ with $\alpha_1=\alpha_2=1/\sqrt2$ (i. e., a minimum uncertainty wave packet centered at 
$q_1=q_2=1$ with mean number of excitations $\langle N(0)\rangle=|\alpha_1|^2+|\alpha_2|^2=1$).
All calculations are done by numerical integration of the time dependent Shrödinger equation in a truncated Fock space for the Hamiltonian (\ref{Hrf})
with the trap and rotation parameters being those in the first row of Table \ref{data_table}.} 
\label{observables_fig}
\end{center}
\end{figure}


%
%
\subsubsection{Mean number of excitations, survival probability.}
We will now consider the mean vibrational number as a function of time in the rotating frame, 
\begin{eqnarray}
 \langle N(t) \rangle &=&\langle \psi(t)|{a}_1^\dag {a}_1+{a}_2^\dag {a}_2|\psi(t)\rangle, 
\end{eqnarray}
where the creation and annhilation operators in each direction are defined in terms of position and momentum operators as usual,
\begin{eqnarray}
 {a}_j&=&\frac{1}{\sqrt2}(\widehat{q}_j+i\widehat{p}_j),\\
{a}_j^\dag&=&\frac{1}{\sqrt2}(\widehat{q}_j-i\widehat{p}_j),
\end{eqnarray}
for $j=1,2$. In the first column of Fig. \ref{observables_fig}, the time evolution of the mean number of excitations during $\pi/2$ rotations designed without final 
excitation using the first row  of Table \ref{data_table} ($n_1=1, n_2=2$) is shown for different initial states:  the ground state of the non-rotating 
trap, an entangled state, and a coherent state.  
Interestingly, Fig. \ref{observables_fig} (first column) demonstrates that the mean excitation
can actually decrease, at least transitorily, with respect to the initial value. Of course  for all states  the final value coincides with  the initial value. 

%
%

The rotation process has been chosen so that the survival probability $P(t)=|\langle \psi(t)|\psi(0)\rangle|^2$ satisfies the condition $P(0)=P(T)$,
due to commensurability. 
In the second column of Fig. \ref{observables_fig}, the probability of finding the system in its initial state is calculated for different initial quantum states.
The revivals are seen clearly in all three cases. The survival of the coherent and entangled states decays at intermediate times much more severely than the one for the ground state. 
Indeed, a classical particle set initially at rest at the bottom of the trap would not be affected  by the trap rotation. 

\begin{figure}[t]
\begin{center}
\includegraphics[width=8cm]{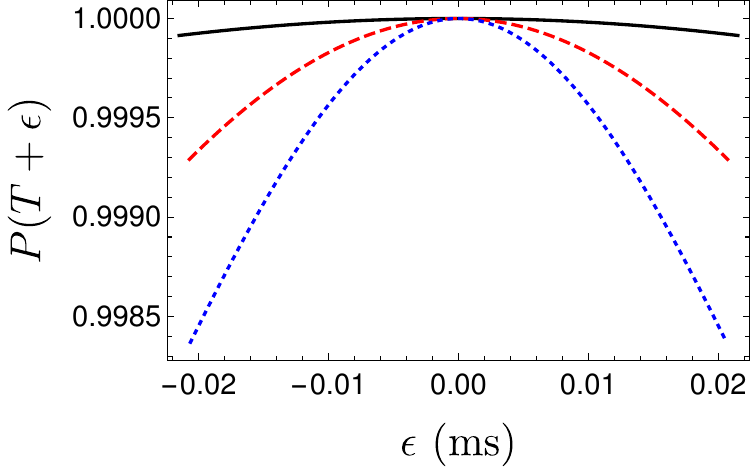}
\caption{(Color online) Stability when the rotation lasts $T+\epsilon$. 
The survival probability is plotted as a function of $\epsilon$ for the example shown in Fig. \ref{observables_fig}d ($\pi/2$ rotation of the ground state)
for $n_1=1$ and different integer values of $n_2$:  $n_2=2$ (black-solid line), $n_2=5$ (red-dashed line), and   $n_2=10$ (blue-dotted line). 
}
\label{stability_fig}
\end{center}
\end{figure}
%

\subsection{Stability}
%
%
As already pointed out in Sec. \ref{fast_rot_sec}, the fastest allowed rotations are  found for 
$n_1=1$ and $n_2\gg1$, which imply very narrow quasi-1D traps with $\omega_2\gg \omega_1$.
However, fast rotations  come with a price, since as $n_2$ increases the ideal result becomes more unstable.  
This can be intuitively understood:  for larger $n_2$ the second normal oscillator oscillates faster  so  it is easier to miss the exact final state due to some small timing error. This is confirmed in Fig. \ref{stability_fig}, which depicts  the survival probability as a function $\epsilon$, a small deviation
from the nominal operation time $T$. For larger $n_2$ the survival becomes less robust. 

This effect can be quantified by approximating the survival probability to second order in $\epsilon$ as
\begin{eqnarray}
P(T+\epsilon)\approx 1-\Delta H^2 \epsilon^2
\end{eqnarray}
with 
\begin{eqnarray}
 \Delta H^2=\langle  \psi(0)|H^2|\psi(0)\rangle - \langle  \psi(0)|H|\psi(0)\rangle^2.
\end{eqnarray}
$\Delta H^2$ depends on the considered initial state as Fig. \ref{observables_fig} (right column) illustrates. 
The survival probability of the ground state, in particular, decays with $\epsilon^2$ at a rate
\begin{eqnarray}
 \Delta H^2=\frac{\dot\theta ^2(\omega_1-\omega_2)^2}{4 \omega_1 \omega_2},
\end{eqnarray}
which, for a given $\omega_1$, increases for faster rotations (larger $\dot\theta$ and $\omega_2$). 

\section{Discussion\label{d}}
Controlling the motion of quantum particles is needed  to manipulate them for fundamental science studies
and to develop different quantum technologies.  In particular, operations which are fast, robust, and do not leave
residual excitations are typically preferred.    
Here we focused on rotating arbitrary states of a single particle in an anisotropic harmonic trap using the rotation speed and rotation time as the only control 
parameters. By ``rotating'' a state here we mean to end at a time $t=T$ with a particle/trap configuration which is identical to the one 
at time $t=0$ but rotated by some angle $\theta_f$ in the laboratory frame.    
As an inverse problem, even such a simple system and operation  involves considerable complexities. 
Since normal modes cannot be found by a point transformation, we have first performed a non point (but canonical) transformation 
to find the normal modes for constant rotation speed. Based on the normal mode analysis  we apply a shortcut-to-adiabaticity protocol in which 
any initial state becomes its rotated version in the final trap. Minimal times are found and a  stability analysis with respect to time errors is performed. 

We may envision several worthwhile and natural extensions of this work such as considering anharmonicities, 
two or more interacting particles in the trap, or, to achieve further flexibility in the rotation times,  
time-dependent rotation speeds $\dot{\theta}(t)$. This time-dependence makes the $A(t)$ matrix in the Hamiltonian of the rotating 
frame time dependent, and following the steps in the main text and Appendix \ref{unitary_transf_app} 
we may perform a time-dependent symplectic transformation 
and find that  the interaction picture effective Hamiltonian will be given by 
$H_I=v^T[S^T(A-\dot G)S]v$, see Eq. (\ref{HI_full_app}).
Finding the time-dependent symplectic transformation $S$ that 
makes the $4\times 4$ matrix $A'=S^T(A-\dot G) S$ diagonal is a challenging open question. 

We cannot fail to point out an analogy between the structure of $A'$ and the effective Hamiltonian used in  superadiabatic  iterations to achieve shortcuts \cite{Demirplak2008,Ibanez2013}. 
If $S$ is set to diagonalize $A$, rather than the whole matrix $A'$,  two uncoupling strategies are: 
to ignore the inertial term ${\cal I}=-S^T \dot G S$  because it is small (this is analogous to an adiabatic approximation), or to compensate it exactly with $-{\cal I}$ (this is analogous to counter-diabatic driving). However  implementing such  
a compensating term is often challenging in practice, in this case it implies crossed operator terms. 
A third route is to apply the next  ``superadiabatic'' iteration, i.e.,  
to find an $S'$ that makes  ${S'}^TA'S'$ diagonal, which produces  a term ${\cal I}'=-S'^T \dot G S'$
in the new Hamiltonian. Further iterations would repeat the same scheme but they do not need to converge so there may be an optimal iteration. Alternatively the coupling term  may be approximated  to achieve convergence \cite{Theis2018}. 
All this is very intriguing and will be  explored elsewhere.

\acknowledgments{We are grateful to K. Takahashi for discussions in early stages of the work. This work was supported by 
the Basque Country Government (Grant No.
IT986-16).}

\appendix

\section{Laboratory frame}
\label{lab_frame_app}
The Hamiltonian for a particle of mass $m$ in a two-dimensional anisotropic harmonic potential with axial frequencies $\omega_1$ and $\omega_2$ 
and with a time varying orientation angle $\theta(t)$ (i. e., which is rotating around the $z$ axis with angular velocity $\dot\theta(t)$)
is  given in laboratory $\{x,y\}$ frame by
\begin{eqnarray}
\label{lab_frame_H}
 H_{\textrm{lab}}&=&\frac{p_x^2}{2m}+\frac{p_y^2}{2m} +\frac{m\omega_1^2}{2} \left[x \cos\theta(t)+y \sin\theta(t)\right]^2\nonumber\\
 &+&\frac{m\omega_2^2}{2} \left[-x \sin\theta(t)+y \cos\theta(t)\right]^2.
\end{eqnarray}
Defining the rotated coordinates and momenta by the relations
\begin{eqnarray}
 \left(\begin{matrix} \tilde q_1\\ \tilde q_2\end{matrix}\right)=R(t)\left(\begin{matrix} x\\y\end{matrix}\right);
 \left(\begin{matrix} \tilde p_1\\ \tilde p_2\end{matrix}\right)=R(t)\left(\begin{matrix} p_x\\p_y\end{matrix}\right)
\end{eqnarray}
with $R(t)$ being the usual rotation matrix
\begin{eqnarray}
\label{rot_matrix}
 R(t)=\left(\begin{matrix} \cos\theta(t)&\sin\theta(t)\\-\sin\theta(t)&\cos\theta(t)\end{matrix}\right),
\end{eqnarray}
the new Hamiltonian is given by 
\begin{eqnarray}
 H&=&\frac{\tilde p_1^2}{2m}+\frac{\tilde p_2^2}{2m} +\frac{1}{2}m\omega_1^2 \tilde q_1^2
 +\frac{1}{2}m\omega_2^2\tilde q_2^2
 -\dot\theta L_z,
\end{eqnarray}
with $L_z= \tilde q_1  \tilde p_2- \tilde q_2 \tilde p_1$.
This last term, which couples coordinates and momenta, accounts for the inertial effects that arise due to the time-dependent 
canonical transformation applied.

\section{Quantum unitary transformations}
\label{unitary_transf_app}
It is also instructive to set  a quantum description by means of a unitary transformation of the Hamiltonian.
As it is well known from group theory, the generators of symplectic matrices are symmetric matrices in the sense that any symplectic matrix $S$ can 
be written in terms of its generator $G$ as $S=e^{2JG}$, $G$ being a symmetric matrix and $J$ the symplectic matrix (\ref{symplectic_Jmatrix}).
Let us define the unitary operator 
\begin{eqnarray}
\label{unitary_operator}
 \mathcal{U}=e^{iv^T Gv},
\end{eqnarray}
where $v^T$ is now regarded as a vector of operators $ v^T=(\widehat{q}_1,\widehat{q}_2,\widehat{p}_1,\widehat{p}_2)$.
The unitarily transformed,  interaction picture Hamiltonian  will be given by
\begin{eqnarray}
\label{HI_U}
 H_I&=&\mathcal{U} H \mathcal{U}^\dag+i \mathcal{\dot U} \mathcal{U}^\dag,
\end{eqnarray}
where the last term arises due to the possible time dependence of the unitary transformation.
For a quadratic  Hamiltonian with the form $H=v^T A v$, see Eq. (\ref{Hrf_matrixform}), and 
the unitary operator $\mathcal{U}$ defined by (\ref{unitary_operator}), it can be shown that the above effective Hamiltonian
is given by
\begin{eqnarray}
\label{HI_full_app}
 H_I&=&v^T\left[S^T\left(A-\dot G\right)S\right]v.
\end{eqnarray}
Details of this calculation are given in  Appendix \ref{detailed_calc_app}.

In a time independent scenario, where $\dot G =\dot S=0$, we have an uncoupled (i. e., without cross terms) 
effective interaction picture Hamiltonian
\begin{eqnarray}
H_I&=&v^T \left(S^T A S\right)v
\end{eqnarray}
since $S^T A S$ is a diagonal matrix as shown in Sec. \ref{symplectic_diag_sec}.
Indeed, the inverse unitary  transformation $\mathcal{U}^\dag (...) \mathcal{U}$ maps all the components $v_j$  to $V_j$,
\begin{eqnarray}
\mathcal{U}^\dag v_j \mathcal{U}=\left(S^{-1} v\right)_j=V_j,
\end{eqnarray}
so that $H$ is recovered,
\begin{eqnarray}
H&=&\mathcal{U}^\dag H_I \mathcal{U}=V^T \left(S^T A S\right)V=v^T A v.
\end{eqnarray}
In summary, the same symplectic transformation that diagonalizes the classical Hamiltonian matrix provides as well a quantum Hamiltonian written as a sum of quadratic operators without cross terms.

If the symplectic transformation $S$ depends on time, the extra term 
$-v^T\left(S^T\dot G S\right)v$ has to be included in the effective 
Hamiltonian to account for the inertial effects. For a time dependent transformation, one would have 
to symplectically diagonalize the full matrix $A-\dot G$.

\section{Detailed calculation of Eq. (\ref{HI_full_app})}
\label{detailed_calc_app}
%
%
Let $\mathcal A$ and $\mathcal B$ be two real symmetric matrices.
Taking into account that the position-momentum commutators $[q_j,p_k]=i\delta_{jk}$ can be summarized as
$[v_j,v_k]=i J_{jk}$, one can find the relation
\begin{eqnarray}
\label{nth_commutator}
 \sum_{n=0}^\infty \frac{[v^T \mathcal Bv,v^T \mathcal A v]_n}{n!}=v^T \left(e^{2i\mathcal B J}\mathcal A e^{-2iJ\mathcal B} \right)v,
\end{eqnarray}
where $[.,.]_n$ denotes the $n$th nested commutator between the involved operators.
Using this result, the two terms in the effective Hamiltonian (\ref{HI_U}) will be calculated separately:

(i) The first  term $\mathcal{U} H \mathcal{U}^\dag$ can be calculated using the Baker-Campbell-Hausdorff (BCH) formula and  the previous result (\ref{nth_commutator}) to sum the series expansion.
For the unitary operator defined in (\ref{unitary_operator}) and a Hamiltonian with the form (\ref{Hrf_matrixform}) we have 
\begin{eqnarray}
 \mathcal{U} H \mathcal{U}^\dag&=&e^{i v^T G v}\left(v^T A v\right)e^{-i v^T G v}=\sum_{n=0}^\infty \frac{[iv^T G v,v^T A v]_n}{n!}\nonumber\\
 &=&v^T \left(e^{-2 G J} A e^{2J G} \right)v=v^T \left(S^T A S \right)v.
\end{eqnarray}

(ii) To calculate the second term $i \mathcal{\dot U} \mathcal{U}^\dag$, 
we must be careful when computing the time derivative of $\mathcal{U}$, since it involves not-commuting operators \cite{Hall2015},
\begin{eqnarray}
 i \mathcal{\dot U} \mathcal{U}^\dag&=&i\sum_{n=0}^\infty \frac{[iv^T G v,iv^T \dot G v]_n}{n!}
 =-v^T \left(e^{-2GJ} \dot G e^{2JG}\right)v\nonumber\\
 &=&-v^T \left(S^T \dot G S\right)v.
\end{eqnarray}
Here, again, Eq.  (\ref{nth_commutator}) has been used to sum the series expansion.

The sum of these two terms leads finally to the interaction picture effective Hamiltonian (\ref{HI_full_app}) 
\begin{eqnarray}
 H_I&=&\mathcal{U} H \mathcal{U}^\dag+i \mathcal{\dot U} \mathcal{U}^\dag=v^T\left[S^T\left(A-\dot G\right)S\right]v.
\end{eqnarray}


\end{document}